\documentstyle[12pt]{article}
\def\be{\begin{equation}}
\def\ee{\end{equation}}
\begin{document}

\hfill{\em In memory of F.A. Berezin} \vglue 1cm 
\centerline{\bf On the functional equation related}
\medskip
\centerline{\bf to the quantum three-body problem\,\footnote
{\,\,Published in Berezin Memorial Volume, 
Amer. Math. Soc. Transl. (2) Vol.{\bf 175}, 15--34, 1996}}

\bigskip
\centerline{V.M. Buchstaber} \smallskip\centerline{\em Department 
of Mathematics and Mechanics}
\smallskip\centerline{\em Moscow State University}\smallskip
\centerline{119899 {\em Moscow, Russia}} \bigskip\centerline{A.M. Perelomov}
\smallskip\centerline{\em Institute of Theoretical and
Experimental Physics,} \smallskip\centerline{117259 {\em Moscow, Russia}} 
\bigskip

\centerline{\bf Abstract} \medskip \noindent In the present paper 
we give the general solution of the functional equation
\[ ( f(x) + g(y) + h(z) )^2 = F(x) + G (y) + H (z),\qquad 
x + y + z = 0 \]
which is related to the exact factorized ground-state wave function for 
the quantum one-dimensional problem of three different particles 
with pairwise interaction.
\vfill\break
\noindent Functional equations connecting several functions and admitting 
a  general analytic solution have 
recently attracted the attention of many mathematicians as  well as 
physicists (for recent results, see, for example, [BC 1990], [BP 1993], 
[BK 1993], [BFV 1994]).

In modern mathematical physics such equations arise in connection with the 
integrable systems of classical and quantum mechanics (see, for example, 
reviews [Pe 1990], [OP 1983]).

In the present paper (a previous version of which appeared as [BP 1993]), 
we investigate one such equation. Namely, we investigate 
the functional equation connecting six unknown functions
\be (f(x) + g(y) + h(z))^2 = F(x) + G(y) + H(z),\qquad x+y+z = 0, \ee
which generalizes the well-known Frobenius--Stickelberger equation 
[FS 1880] and is related to the exact factorized ground-state wave 
function for the quantum one-dimensional problem of three 
different particles with pairwise interaction. We give the general 
non-degenerate solution of this equation.

{\bf 1.} Let us recall first that an analogous (but simpler) 
equation  for the special case of three identical particles was 
considered earlier by B. Sutherland [Su 1975] and F. Calogero [Ca 1975]. 
Namely, in the paper [Su 1975] the one-dimensional 
many-body problem of $n$ identical particles with pair interaction 
was considered, whose exact ground-state wave function 
$\Psi_0(x_1, x_2,\ldots ,x_n)$ is factorized
\be 
\Psi_0 (x_1,x_2,\ldots ,x_n) = \prod_{j<k} \psi(x_j - x_k). \ee
It was shown that the logarithmic derivative of $\psi (x)$
\be 
f(x) = \psi'(x)/ \psi (x) \ee
must satisfy the functional equation
\be \begin{array}{l}
f(x)\,f(y) + f(y)\,f(z) + f(z)\,f(x) = F(x) + F(y) + F(z),\\
x + y + z = 0,\end{array} \ee
where $f(x)$ ($F(x)$, respectively) is an odd (even) function
\be 
f(-x) = -\,f(x),\qquad F(-x) = F(x). \ee
In [Su 1975], a partial solution of equations (4) and (5) was also found.

The general solution of equations (4) and (5) was found in [Ca 1975] 
(for review of this and the related problems, see also [OP 1983]). 
This solution has the form
\be
f(x) = \alpha\,\zeta (x; g_2, g_3 ) + \beta x, \ee
where $\zeta (x)$ is the Weierstrass zeta-function (see, for instance, 
[WW 1927]).

In the present paper, we consider only the three-body problem but 
in the general case  when all three particles are different from each other.

In this case, the ground-state wave function has the form 
\be
\Psi_0 (x_1, x_2, x_3) = \psi_1 (x_2 - x_3)\,\psi_2 (x_3 - x_1)\,
\psi_3 (x_1 - x_2) \ee
and satisfies the Schr\"odinger equation
\begin{eqnarray}
&&-\Delta \psi_0 + U \psi_0 = E_0 \Psi_0, \\
&&U= u_1 (x_2 - x_3) + u_2 (x_3 - x_1 ) + u_3 (x_1 - x_2 ). \end{eqnarray}
Substituting $\Psi_0$ from (7) into (8), we obtain
\begin{eqnarray}
\Psi_0^{-1}\,\Delta \Psi_0 &=& U - E_0 = 3\left( f_1^2(x_2 - x_3) +
f_2^2(x_3 - x_1 ) + f_3^2(x_1 - x_2 )\right) \nonumber \\
&&- \left( f_1(x_2 - x_3 ) + f_2(x_3 - x_1) + f_3(x_1 - x_2)\right) ^2 
\nonumber \\
&&+2\left( f'_1 (x_2 - x_3 ) + f'_2 (x_3 - x_1 ) + f'_3 (x_1 - x_2)\right) ;\\
f_j &=& {\psi'_j / \psi_j}.\nonumber \end{eqnarray}
Hence, for the potential energy $U(x_1, x_2, x_3)$ to have  the form
of pairwise interactions (9), three functions
\be 
f(x) = f_1 (x),\qquad g(y) = f_2 (y),\qquad h(z) = f_3(z)\ee
must satisfy the functional equation
\be
\left( f(x) + g(y) + h(z)\right) ^2 = F(x) + G(y) + H(z),\qquad 
x + y + z = 0. \ee

From (10) - (12) it results the following expression for the potential 
energies:
\begin{eqnarray}
u_1 (x) &=& 3\,f^2 (x) + 2\,f'(x) - F(x) + \varepsilon_1, \nonumber \\
u_2(x)&=& 3\,g^2 (x) + 2\,g'(x) -G(x) + \varepsilon_2, \\
u_3(x) &=& 3\,h^2 (x) + 2\,h'(x) - H(x) + \varepsilon_3,\nonumber \\
&&\varepsilon_1 +\varepsilon_2 +\varepsilon_3 = E_0 \nonumber 
\end{eqnarray}
\vfill\break

{\bf 2.} Let us consider the meromorphic solutions of the equation
\be
\left( f(x)+g(y)+h(z)\right) ^{2}=F(x)+G(y)+H(z)\ee
satisfying the condition $x+y+z = 0$.

Let us call the solution of equation (14) {\em nondegenerate} if the 
functions $f(x)$, $g(x)$ and $h(x)$ have the pole in a finite domain 
of complex $x$-plane.

The main result of this paper is the following

{\bf Theorem}. {\em The general nondegenerate solution of equation} 
(14) {\em in the class of meromorphic functions has the form}
\begin{eqnarray}
f(x) &=& \alpha \zeta (x-a_1; g_2, g_3) + \beta x + \gamma_1 ,\\
g(x) &=& \alpha \zeta (x-a_2; g_2, g_3 ) + \beta x + \gamma_2, \\
h(x) &=& \alpha \zeta (x-a_3; g_2, g_3 ) + \beta x + \gamma_3,\\
F(x) &=& \alpha^2 {\cal P} (x-a_1; g_2, g_3) + 2\gamma \alpha \zeta (x-a_1;
g_2, g_3) + \frac{\gamma^2}3,\\
G(x) &=& \alpha^2 {\cal P} (x-a_2; g_2, g_3) + 2\gamma \alpha \zeta (x-a_2;
g_2, g_3) + \frac{\gamma^2}3,\\
H(x) &=& \alpha^2 {\cal P} (x-a_3; g_2, g_3) + 2\gamma \alpha \zeta (x-a_3;
g_2, g_3) + \frac{\gamma^2}3, \end{eqnarray}
{\em where}
\be 
a_1 + a_2 + a_3 =0,\qquad \gamma _1+\gamma _2+\gamma _3=\gamma . \ee

{\bf Proof .} The proof of the theorem is divided on several steps.

Let us begin with

{\bf Lemma 1}. {\em The functions ($f(x)$, $g(y)$, $h(z)$) satisfy  
equation} (14) {\em for the corresponding functions} ($F(x)$, $G(y)$, $H(z)$) 
{\em if and only if the equation}
\be 
\mbox{det}\,\left( \begin{array}{ccc}
f''(x)&g''(y)&h''(z)\\
f'(x) &g'(y)& h'(z)\\ 1&1&1\end{array} \right) =0 \ee
{\em can be solved under condition} $x+y+z = 0$.

{\bf Proof.} Let us apply to equation (1) the operator
\[ 
\partial _{-} \cdot \frac{\partial }{\partial y}\cdot \frac{\partial }
{\partial x}, \]
where 
\[ 
\partial _{-} =\frac{\partial }{\partial x} - \frac{\partial }{\partial y}.\]
This gives:
\begin{eqnarray}
\frac{\partial }{\partial x} &\colon & 2\,(f'(x)-h'(z))(f(x)+g(y)+h(z)) =
F'(x) - H'(z),\\
\frac{\partial }{\partial y}\,\frac{\partial }{\partial x} &\colon & 
2\,h''(z)\,(f(x)+g(y)+h(z)) \nonumber \\
&+& 2\,(f'(x)-h'(z))(g'(y)-h'(z)) = H''(z),\\
{\partial _{-}}\frac{\partial }{\partial y}\,\frac{\partial }{\partial x} 
&\colon & h''(z)(f'(x)-g'(y))\nonumber \\
&+&f''(x)\,(g'(y)-h'(z))+g''(y)\,(h'(z)-f'(x)) = 0.\end{eqnarray}

Here we use the fact that $\partial _{-}$ is a differential operator 
and that $\partial _{-}h'(z) = \partial _{-} h''(z) = 0$.
Hence, if functions $f(x)$, $g(y)$, $h(z)$ satisfy equation (1), then 
these functions also satisfy equation (25) which can be obviously rewritten 
in the form (22).

Conversely, let the functions $f(x)$, $g(y)$, $h(z)$ satisfy (22) and, 
consequently, (25). Equation (25) may be rewritten as
\[ \partial _{-}\,[h''(z)(f(x)+g(y)+h(z)) + (f'(x)-h'(z))(g'(y)-h'(z))] = 0.
\]
Then there is the function $H_{1}(z)$ satisfying the following equation:
\be
h''(z)\,(f(x)+g(y)+h(z))+(f'(x)-h'(z))(g'(y)-h'(z)) = H_{1}(z)\ee

Let us note that equation (26) is equivalent to the equation
\[
\frac{\partial }{\partial y}\,[(f'(x)-h'(z))(f(x)+g(y)+h(z))] = H_{1}(z). \]
Therefore, there are functions $F_{1}(x)$ and $H_{2}(z)$ such that $H_{2}'(z) 
= H_{1}(z)$, and
\be
(f'(x)-h'(z))(f(x)+g(y)+h(z)) = F_{1}(x) - H_{2}(z).\ee 

On the another hand, equation (27) is equivalent to
\[
\frac{\partial }{\partial x}\,(f(x)+g(y)+h(z))^{2} = 2(F_{1}(x)-H_{2}(z)),\]
i.e., there are functions $F(x)$, $G(y)$ and $H(z)$ such that  $F'(x) =
2F_{1}(x)$,  $H'(z) = 2H_{2}(z)$, and
\[
(f(x)+g(y)+h(z))^{2} = F(x)+G(y)+H(z).\]

Thus, Lemma 1 has been proved.

{\bf Lemma 2. }{\em  Equation} (14) {\em is invariant under the following 
transformations:}
\begin{eqnarray*}
f(x) &\to & f_{0}+a_{1}x+a_{2}f(a_{3}x+\alpha _{1}),\\
F(x) &\to & F_{0}+a_{4}x+a_{2}^{2}F(a_{3}x+\alpha _{1}) + 2a_{2}c f(a_{3}x+
\alpha _{1}),\\
g(y) &\to & g_{0}+a_{1}y+a_{2}g(a_{3}y+\alpha _{2}), \\
G(y) &\to & G_{0}+a_{4}y+a_{2}^{2}G(a_{3}y+\alpha _{2}) + 2a_{2}cg(a_{3}y+
\alpha _{2}),\\
h(z) &\to & h_{0}+a_{1}z+a_{2}h(a_{3}z+\alpha _{3}),\\
H(z) &\to & H_{0}+a_{4}z+a_{2}^{2}H(a_{3}z+\alpha _{3}) + 2a_{2}c\,
h(a_{3}z+\alpha _{3}),\end{eqnarray*}
{\em where} $a_{k} (k=1,\ldots ,4$) {\em and} $c$ {\em are free parameters}
\be 
f_{0}+g_{0}+h_{0} = c,\qquad F_{0}+G_{0}+H_{0} = c^{2},\qquad \alpha _{1}+
\alpha _{2}+\alpha _{3} = 0.\ee

This Lemma is proved by a direct calculation.

{\bf Corollary 3.} {\em For appropriate values of the parameters}
$\alpha _{1}$, $\alpha _{2}$, $\alpha _{3}$, {\em all the functions } 
$(f(x), g(y), h(z))$, $(F(x), G(y), H(z))$ {\em are regular at} $x=0$, 
$y=0$, $z=0$, {\em respectively.}

The proof follows from the fact that the set of poles of the meromorphic 
function of one complex variable is discrete. Thus, we may 
suppose that all the functions are regular at $x=0, y=0, z=0$.

{\bf Definition 4.}  Let us call the solution of equation (14) {\em totally 
degenerate} if at least one of functions $f(x), g(x),$ and $h(x)$ 
is linear.

The next Lemma describes all the degenerate solutions of equation (14).

{\bf Lemma 5.} {\em Let} $(f(x), g(y), h(z))$, $(F(x), G(y), H(z))$ 
{\em be a totally degenerate solution of  equation} (14).

Three cases are possible.
\begin{description}
\item[1.]  All three functions $f(x)$, $g(y)$ and $h(z)$  are linear. Then 
\[ \begin{array}{l}
f(x)=f_{0}+f_{1}x,\qquad F(x)=F_{0}+F_{1}x+(f_{1}-g_{1})(f_{1}-h_{1})x^{2},\\
g(y)=g_{0}+g_{1}y,\qquad G(y)=G_{0}+G_{1}y+(g_{1}-f_{1})(g_{1}-h_{1})y^{2},\\
h(z)=h_{0}+h_{1}z,\qquad H(z)=H_{0}+H_{1}z+(h_{1}-g_{1})(h_{1}- f_{1})z^{2}.
\end{array} \]
Here $f_{0}, f_{1}, g_{0}, g_{1},h_{0},h_{1}$ are free parameters.

Let $f_{0}+g_{0}+h_{0} = c$. Then
\[
F_{0}+G_{0}+H_{0} = c^{2},\qquad F_{1} = b+2cf_{1},\qquad G_{1} = b+2cg_{1},
\qquad H_{1} = b+2ch_{1}, \]
and $b$ is a free parameter.
\item[2.] Two of the functions $f(x)$, $g(y)$ and $h(z)$ are linear. 
For example, 
\[
g(y) = g_{0}+g_{1}y,\qquad h(z) = h_{0}+h_{1}z.\]
Then $f(x)$ is an arbitrary function and  
\[ g(y) = g_{0}+ay,\qquad h(z) = h_{0}+az,\qquad G(y) = G_{0}+by,
\qquad H(z) = H_{0}+bz, \]
\[ F(x) = [g_{0}+h_{0}-ax+f(x)]^{2} - (G_{0}+H_{0}-bx). \]
Here $g_{0}, h_{0}, a, b, G_{0}, H_{0}$ are free parameters.
\item[3.] Only one of the functions $f(x)$, $g(y)$ and $h(z)$ is linear. 
For example, $h(z) = h_{0}+h_{1}z$. Then
\[ \begin{array}{l}
f(x) = f_{0}+ax+c_{1}\exp (\lambda x), \\
F(x) = F_{0}+bx+c_{1}\exp (\lambda x)\,(2c+c_{1}\exp (\lambda x)),\\
g(y) = g_{0}+ay+c_{2}\exp (\lambda y), \\
G(y) = G_{0}+by+c_{2}\exp (\lambda y)\,(2c+c_{2}\exp (\lambda y)),\\
h(z) = h_{0}+az, \\
H(z) = H_{0}+bz+2c_{1}c_{2}\exp (-\lambda z). \end{array} \]
Here $a, b, c, c_{1}, c_{2}, \lambda $ are  free parameters, and
\[ 
f_{0}+g_{0}+h_{0} = c,\qquad F_{0}+G_{0}+H_{0} = c^{2}. \]
\end{description}

{\bf Proof.}

{\bf Case 1.} From (22) it follows that $f(x)$,  $g(y)$ and $h(z)$ are the 
arbitrary linear functions. The form of the functions $F(x)$, $G(y)$ and 
$H(z)$ can be reconstructed directly from (14) by taking into account 
the identity $2\,xy=z^{2}-x^{2}-y^{2}$.

{\bf Case 2.} From (22), we obtain
\[ f''(x)(g_{1}-h_{1}) = 0. \]
If $f''(x) \neq 0$, then $g_{1}=h_{1}$, and $f(x)$ is the arbitrary function. 
The form of the functions $F(x), G(y), H(z)$ can be reconstructed immediately.

{\bf Case 3.} From (24), we get
\[ 2(f'(x)-h_{1})(g'(y)-h_{1}) = H''(-x-y).\]

If $f'(x)$ and $g'(y)$ are not constants, then according to the 
classical Cauchy--Pexider result [Ca 1821] (see also [Ab 1823]), we obtain 
\[ 
f'(x)-h_{1} = \tilde {c}_{1}\,\exp (\lambda x),\qquad g'(y)-h_{1} =
\tilde {c}_{2}\,\exp (\lambda x),\]
where $\tilde {c}_{1}, \tilde {c}_{2}$ and $\lambda $ are free parameters.
Therefore,
\[ f(x) = f_{0}+h_{1}x+c_{1}\exp (\lambda x),\qquad g(y) = g_{0}+h_{1}y+c_{2}
\exp (\lambda y),\]
where $c_{k} = {\tilde {c}_{k}/\lambda }$, $k=1, 2$. 
The form of the functions $F(x),G(y), H(z)$ can be reconstructed easily.
Lemma is proved.

The functions $f(x), g(x), h(x)$ from equation (1) will be regarded as 
nondegenerate solutions of equation (1).

{\bf Lemma 6.} {\em For the appropriate values of the parameters} 
$f_{0}, g_{0}, h_{0}, a_{1}, F_{0}, G_{0}$ ({\em see Lemma} 2) {\em we have} 
\be
f(0) = g(0) = h(0) = 0,\qquad h'(0) = 0,\qquad F(0) = G(0).\ee

The proof is easy.

{\bf Lemma 7.} {\em For an appropriate choice of the parameters } 
$\alpha _{1}$ {\em and} $\alpha _{2}$, {\em we have the relation} 
$f(x) \neq g(x)$.

{\bf Proof.} Suppose on the contrary that
\be
f(x+\alpha _{1})-f(\alpha _{1}) \equiv g(x+\alpha _{1})-g(\alpha _{2}) \ee 
for all $\alpha _{1}$ and $\alpha _{2}$ in any neighborhood of the point 
$x=0$. Differentiating (30), we obtain
\[
\frac{\partial f(x+\alpha _{1})}{\partial x} = \frac{\partial f(x+
\alpha _{1})}{\partial \alpha _{1}} = f'(\alpha _{1}), \]
i.e., 
\[
f(x+\alpha _{1}) = f'(\alpha _{1})x+f(\alpha _{1}).\]
contradicting the assumption that the solution is nondegenerate. 
Lemma is proved.

Hence, it is sufficient to find all the nondegenerate solutions 
of equation (1) under the following additional conditions: $f(x) \neq 
g(x)$ and $f(0) = g(0) = h(0)$,  $h'(0) = 0$, $F(0) = G(0) =0$.

Interchanging $x$ and $y$ in equation (14), we obtain
\be 
(f(y)+g(x)+h(z))^{2} = F(y)+G(x)+H(z).\ee
Subtracting (31) from (14), we see that
\begin{eqnarray*} 
&& [(f(x)-g(x))-(f(y)-g(y))]\,[(f(x)+g(x))+(f(y)+g(y)+2h(z))]\\
&=& (F(x)-G(x))-(F(y)-G(y)). \end{eqnarray*}
The last equation can be rewritten as 
\be
\varphi (x+y) = \eta (x)+\eta (y)-\frac{\gamma (x)-\gamma (y)}{\xi (x)-
\xi (y)},\ee 
where $\varphi (x)=-\,2h(-x)$, $\eta (x)=f(x)+g(x)$, $\xi (x)=f(x)-g(x)$, 
$\gamma (x)=F(x)-G(x)$, and  $\varphi (0) = \varphi '(0) = \eta (0)-
\gamma (0)=\xi (0) = 0$, and $\varphi ''(x) \neq 0.$

{\bf Definition 8.}  Let us call the solution $(\varphi , \eta , \xi ,
\gamma )$ of equation (32)  {\em normalized} if the following 
 initial conditions are satisfied:
\[
\xi '(0) = 1,\qquad \eta '(0)=0.\]

{\bf Lemma 9}. {\em The map}
\be
(\varphi ,\eta , \xi ,\gamma ) \to \left( \varphi ,\,\eta +b_{1}\xi ,\,
b_{2}\xi ,\,b_{2}\left( \gamma +b_{1}\xi ^{2}\right) \right),\ee 
{\em where} $b_{1}$ {\em and} $b_{2}$ {\em are parameters and}, 
$b_{2} \neq 0$, {\em defines a group action. Each orbit of this group 
contains one and only one solution}.

{\bf Proof.} The first statement may be checked by a direct computation. 
To prove the second one, let us differentiate equation (32) with respect 
to $y$. At the point $y=0$ we have
\[
\varphi '(x) = \eta '(0)+\frac{\gamma '(0)}{\xi (x)}-\xi '(0)\,\frac
{\gamma (x)}{\xi (x)^2}. \]
Assuming $\varphi (x)$ to be regular at $x=0$ and $\varphi ''(x) \neq 0$, 
it is easy to check that $\xi '(0) \neq 0$.

Applying the transformation (33) with $b_{2} = (\xi '(0))^{-1}$, 
$b_{1} = -\eta '(0)/\xi '(0)$ to the solution $(\varphi , \eta , \xi ,
\gamma )$, we obtain a normalized solution. Lemma is proved.

In what follows, the solutions are assumed to be normalized, unless the 
contrary is asserted. Let us now consider the functional equation
\be \begin{array}{l}
\varphi (x+y) = \varphi (x)+\varphi (y)+\tau (x)\tau (y)A(x+y),\nonumber \\
\varphi (0) = \varphi '(0) = \tau (0) = \tau ''(0) = 0,\qquad \tau '(0) =1.
\nonumber \end{array} \ee

{\bf Lemma 10.} {\em For any solution} $(\varphi , \eta , \xi , \gamma)$ 
{\em of equation} (32), {\em there is an unique solution} 
$(\varphi , \tau , A)$ {\em of equation} (34) {\em such that}
\begin{eqnarray}
\xi (x)&=& {\tau (x)\over {\tau '(x)-b_{3}\tau (x)}},\\
\eta (x) &=& \varphi (x) - \varphi '(x)\xi (x),\\
\gamma (x) &=& -\varphi '(x)\xi (x)^{2},\end{eqnarray}
{\em where} $b_{3} = \xi ''(0)$ {\em is a free parameter.}

{\bf Proof.} Let $(\varphi ,\tau ,A)$ be a solution of  equation (34). 
Then acting on (34) by the operator $\partial _{-} = ({\partial }/
{\partial x}-{\partial }/{\partial y})$, we obtain
\[
0 = \varphi '(x)-\varphi '(y)+(\tau '(x)\tau (y) - \tau (x)\tau '(y))
A(x+y),\]
i.e.,
\be 
A(x+y) = -\,\frac{\varphi '(x) -\varphi '(y)}{\tau '(x)\,\tau (y)-\tau (x)\,
\tau '(y)}\,.\ee
Hence, we can transform  equation (34) to the equation
\be 
\varphi (x+y) = \varphi (x)+\varphi (y)+\tau (x)\tau (y)\,\frac
{\varphi '(x)-\varphi '(y)}{\tau (x)\tau '(y)-\tau '(x)\tau (y)}\,.\ee

On the other hand,
\begin{eqnarray*}
\frac{\tau (x)\tau (y)}{\tau (x)\tau '(y)-\tau '(x)\tau (y)} &=&
\frac{\tau (x)}{\tau '(x)}\,\frac{\tau (y)}{\tau '(y)}\,\frac1{(\tau (x)/
\tau '(x)-b_{3})-(\tau (y)/\tau '(y)-b_{3})}\\
&=& \frac{\xi (x)\,\xi (y)}{\xi (x)-\xi (y)},\end{eqnarray*}
where the function $\xi (x)$ may be expressed in terms of $\tau (x)$ by 
formula (35) with the free parameter $b_{3}$. Therefore,
\[ 
\varphi (x+y) = \varphi (x)+\varphi (y)+\xi (x)\xi (y)\,\frac{\varphi '(x)-
\varphi '(y)}{\xi (x)-\xi (y)}. \]

Substituting the expressions for $\eta (x)$ and $\gamma (x)$ from (36) 
and (37), we obtain the solution $(\varphi , \eta , \xi , \gamma )$ of 
equation (32).

Now let $(\varphi , \eta , \xi , \gamma )$ be a solution of equation (32). 
Substituting $y=0$ in (32), we obtain 
\[
\varphi (x) = \eta (x)-{\gamma (x)\over \xi (x)},\]
i.e., $\gamma (x) = \xi (x)\delta (x)$, where $\delta (x) = \eta (x)-
\varphi (x)$, and  our initial conditions $\varphi '(0) = \eta '(0)=0=
\varphi (0) = \eta (0)$ are satisfied.

Hence, $\gamma '(0)=0$, and from the formula for $\varphi '(x)$ obtained in 
the proof of Lemma 9, we obtain
\[
\gamma (x) = -\varphi '(x)\xi ^{2}(x),\qquad \eta (x) = 
\varphi (x)-\varphi '(x)\xi (x),\]
as asserted in (36) and (37). Let us note that formula (35) may be 
regarded as the differential equation for the function $\tau (x)$. Solving 
this equation with initial conditions $\tau (0)=0, \tau '(0)=1$ we obtain the 
function $\tau (x)$. If, moreover, we take $b_{3}=\xi ''(0)$, this function 
will  satisfy the condition ${\tau}''(0)=0.$

Substituting now the expressions for $\xi (x),\eta (x),\gamma (x)$ into 
equation (32), we obtain equation (39).

Let us apply the operator $\partial _{-}$ to equation (39). We obtain 
\[
\partial _{-}\left( \frac{\varphi '(x)-\varphi '(y)}{\tau (x)\,
\tau '(y)-\tau '(x)\,\tau (y)}\right)  \equiv 0. \]
Thus, we have proved that the functions $\varphi (x)$ and $\tau (x)$ 
determine the function $A(x)$ given by  expression (38). Lemma is 
proved.

So, we have shown how  to construct all the solutions of equation (32) 
using the solutions of equation (34).

Now we describe the general analytical solution of equation (14).

{\bf Lemma 11.} {\em Let} $(\varphi , \tau ,A)$ {\em be a solution of 
equation} (34). ({\em Let us recall that} $\varphi (0) = \varphi '(0) = 
\tau (0) = \tau ''(0)= 0$ {\em and} $\tau '(0) = 1$.) {\em Then the function} 
$u(x) = \varphi '(x)$ {\em is a solution of the equation}
\be 
(u')^{2} = c_{3}u^{3}+4c_{2}u^{2}+2c_{1}u+c_{0}^{2},\qquad 
u(0) = 0,\quad u'(0) = c_{0},\ee
{\em and if} $c_{0} = 0$, {\em then} $c_{1} \neq 0$.

{\em The functions} $\tau (x)$ {\em and} $A(x)$ {\em satisfy the following 
equations:}
\begin{eqnarray}
\frac{\tau '(x)}{\tau (x)} &=& \frac12\,\frac{u'(x)+c_{0}}{u(x)}\,,\\
\frac{A'(x)}{A(x)} &=& \frac12\,\frac{u'(x)-c_{0}}{u(x)}.\end{eqnarray}
{\em If} $c_{0}=0$, {\em then} $u(x) = \frac12\,c_{1}\,\tau (x)^{2}$, 
{\em and} $A(x) = \frac12\,c_1\,\tau (x)$.

{\bf Proof. } Let us consider the first three derivatives with respect to $y$ 
of equation (34)
\begin{eqnarray*}
\varphi '(x+y) &=& \varphi '(y)+\tau (x)[\tau '(y)A(x+y)+\tau (y)A'(x+y)],\\
\varphi ''(x+y) &=& \varphi ''(y)+\tau (x)[\tau ''(y)A(x+y)+2\tau '(y)A'(x+y)
\\
&&+\tau (y)A''(x+y)],\\
\varphi '''(x+y)&=& \varphi '''(y)+\tau (x)[\tau '''(y)A(x+y)+3\tau ''(y)
A'(x+y)\\
&+&3\tau '(y)A''(x+y)+\tau (y)A'''(x+y)].\end{eqnarray*}

Taking $y=0$ and making use of the initial conditions for $\varphi (x)$ and
$\tau (x)$, we get
\begin{eqnarray}
\varphi '(x) &=& \tau (x)\,A(x),\\
\varphi ''(x)&=& \varphi ''(0)+2\tau (x)\,A'(x),\\
\varphi '''(x) &=& \varphi '''(0)+\tau (x)\,[\tau '''(0)\,A(x)+3A''(x)].
\end{eqnarray}
Let  $\varphi _{k} = \varphi ^{(k)}(0)$ and $\tau _{3} = \tau '''(0)$.
From (43) and (44) we obtain
\be 
\frac{\varphi ''(x)-\varphi _{2}}{\varphi '(x)} = 2\,\frac{A'(x)}
{A(x)}.\ee
From (45) and (43) it follows that
\be
\frac{\varphi '''(x)-\varphi _{3}}{\varphi '(x)} =\frac{\tau _{3}\,A(x)+
3A''(x)}{A(x)}.\ee 

Making use of the identity
\[
\frac{A''}{A} = \left( \frac{A'}{A}\right) ' + \left( \frac{A'}{A}\right) ^
{2} \]
for the quantity $\varphi '(x) = u(x)$, we obtain the following equation 
(see equations (46) and (47)):
\[
\frac{u''-\varphi _{3}}{u} = \tau _{3}+3\left( \frac12\,\frac{u'-\varphi _{2}}
{u}\right) ' + \frac34\left( \frac{u'-\varphi _{2}}{u}\right) ^{2}. \]
This equation may be rewritten as follows:
\be \begin{array}{l}
4(u''-\varphi _{3})u = 4\,\tau _{3}u^{2} + 6\,[uu''-u'(u'-\varphi _{2})] +
3\,( u'-\varphi _{2})^{2},\nonumber \\
2\,uu''-3(u')^{2}+4\tau _{3}u^{2} + 4\varphi _{3}u + 3\varphi _{2}^{2} = 0.
\nonumber \end{array} \ee

Let
\[
\tau _{3} = c_{2},\qquad \varphi _{3}=c_{1},\qquad \varphi _{2}=c_{0}.\]
Equation (48) admits the integrating factor $u^{-4}\,u'$ and may be reduced 
to the following equation
\be
(u^{-3}(u')^{2})' = 4c_{2}(u^{-1})' + 2c_{1}(u^{-2})' + c_{0}^{2}(u^{-3})'.\ee
Integrating (49) and multiplying the result by $u^{3}$, we obtain equation 
(40), where $c_{3}$ is the integration constant. Equation (42) follows 
from (46). Then from equation (43) we obtain:
\[
u'(x) = \tau '(x)A(x) + \tau (x)A'(x).\]
From (44) it follows that
\[
\tau (x)\,A'(x) = \frac12\,(u'(x)-c_{0}). \]
Making use of this fact, we obtain
\[
\tau '(x)A(x) =\frac12\,(u'(x)+c_{0}). \]
Dividing this equation by equation (43), we come to equation (41). 
Note that if $c_{0}=0$, equations (41),(42), and conditions $\tau (0)=0, 
\tau '(0)=1$ imply
\[
u(x) = \frac{c_{1}}{2}\,\tau (x)^{2}, \qquad A(x) = \frac{c_{1}}{2}\,
\tau (x). \]
In particular, it follows that $c_{1}\neq 0$ if $c_{0}=0$. 
Lemma is proved.

Consider the Weierstrass function $\wp (x)$ with parameters $g_{2}$ and 
$g_{3}$. We have
\[ 
\wp '(x)^{2} = 4\,\wp (x)^{3}-g_{2}\,\wp (x)-g_{3}.\]

{\bf Lemma 12.} {\em The general solution of  equation} (40) {\em may be 
written in one of the following equivalent forms:}
\begin{eqnarray}
u(x) &=& \frac{4}{c_{3}}\,(\wp (x+\alpha )\cdot \wp (\alpha )),\\
u(x) &=& c_{1}\psi (x) + \frac{c_{0}^{2}\,c_{3}}{2}\,\psi (x)^{2} + 
c_{0}\,\psi '(x),\end{eqnarray}
{\em where}
\be
\psi (x) = \frac12\,\frac1{\wp (x) -c_{2}/3}\,. \ee
{\em Here} $\wp (x)$ {\em is the Weierstrass function with parameters}
\be 
g_{2} = 3\left( \frac{2c_{2}}3\right) ^{2}-\frac{c_{1}c_{3}}{2},\qquad 
g_{3} =-\left( \frac{2c_{2}}{3}\right) ^{3}+\frac{c_{1}c_{2}c_{3}}{6}
-\left( \frac{c_{0}c_{3}}{4}\right) ^{2},\ee
{\em and}
\[
\wp (\alpha ) = \frac13\,{c_{2}},\qquad \wp '(\alpha ) = \frac14\,
{c_{0}c_{3}}.\]

{\bf Proof.} Formula (50) gives:
\[
(u'(x))^{2} =\frac{16}{c_{3}^{2}}\,\left[ 4\wp (x+\alpha )^{3}-g_{2}\wp (x+
\alpha )-g_{3}\right].\]

On the other hand,
\begin{eqnarray*}
(u'(x))^{2} &=& c_{3}\left[ \frac4{c_{3}}\,(\wp (x+\alpha )-\wp (\alpha ))
\right] ^{3}\\
&+& 4c_{2}\left[ \frac4{c_{3}}\,(\wp (x+\alpha )-\wp (\alpha )\right] ^{2} 
+ 2c_{1}\left[ \frac4{c_{3}}\,(\wp (x+\alpha )-\wp (\alpha ))\right] 
+ c_{0}^{2}.\end{eqnarray*}
Hence,
\[ \begin{array}{l}
16\left[ 4\wp (x+\alpha )^{3}-g_{2}\wp (x+\alpha )-g_{3}\right] \\
=4^{3}\left[ \wp (x+\alpha ) -\wp (\alpha )\right] ^{3} + 4^{3}c_{2}
\left[ \wp (x+\alpha )-\wp(\alpha )\right] ^{2}\\
 + 8c_{1}c_{3}\,[\wp (x+\alpha )-\wp (\alpha )] +c_{0}^{2}c_{3}^{2}.
 \end{array} \]

Let us compare the coefficients of terms of the same degree in 
$\wp (x+\alpha )$. This shows that formula (50) with parameters $g_{2}$ and 
$g_{3}$ follows from (53). To deduce (51) from (50) one makes use of the 
addition theorem for the $\wp $-function (see, e.g., [WW 1927]).
\[ 
\wp (x+\alpha )-\wp (\alpha ) = -(\wp (x)+2\wp (\alpha ))+\frac14
\left( \frac{\wp '(x)-\wp '(\alpha )}{{\wp (x)-\wp(\alpha )}}\right) ^{2}.\]
Therefore,
\begin{eqnarray*}
&&(\wp (x+\alpha )-\wp (\alpha ))\left( \wp (x)-\wp (\alpha )\right) ^{2}\\ 
&=&-\,(\wp (x)+2\wp (\alpha ))\left( \wp (x)^{2}-2\wp (x)\wp (\alpha )+
\wp (\alpha )^{2}\right) \\
&+&\frac14\left( 4\wp (x)^{3}-g_{2}\wp (x)-g_{3}-2\wp '(x)\wp '(\alpha )+
\wp '(\alpha )^{2}\right)\\
&=&3\wp (x)\,\wp (\alpha )^{2}-2\wp (\alpha )^{3}-\frac{g_{2}}4\,\wp (x)-
\frac14\,g_{3}-\frac12\,\wp '(x)\,\wp '(\alpha )+\left( \frac
{\wp '(\alpha )}{2}\right) ^{2}\\
&=&\left( 3\wp (\alpha )^{2}-\frac14\,g_{2}\right) (\wp (x)-\wp (\alpha ))-
\frac12\,\wp '(x)\wp '(\alpha ) + \frac12\,{\wp '(\alpha )^{2}}.
\end{eqnarray*}
Hence,
\begin{eqnarray}
\wp (x+\alpha )-\wp (\alpha )&=&\frac12\,\frac{\wp '(x)}{(\wp (x)-
\wp (\alpha ))^{2}}\,\wp '(\alpha )\nonumber \\
&+& \frac{3\wp (\alpha )^{2}-g_{2}/4}{\wp (x)-\wp (\alpha )} + \frac12\,
\left( {\wp '(\alpha )}{\wp (x)-\wp (\alpha )}\right) ^{2}.\end{eqnarray}
This gives:
\[
\wp '(\alpha ) = \frac14\,c_{0}c_{3},\qquad 3\,\wp (\alpha )^{2}-\frac14\,
g_{2} = \frac18\,c_1c_3. \]

Formula (51) follows from equation (54) by dividing by $c_{3}/4$. 
Lemma is proved.

{\bf Corollary 13.} {\em The general solution of equation} (40) {\em has 
the form}
\be
u_{*}(x) = c_{1}\left( \frac{\cosh 2\sqrt {c_{2}}x-1}{(2\sqrt {c_{2}})^{2}}
\right) +c_{0}\,\frac{\sinh 2\sqrt {c_{2}}x}{2\sqrt {c_{2}}} \ee
{\em as} $c_{3}\to 0$.

{\bf Proof.} Let
\[
u_{*}(x) = \lim _{c_{3}\to 1}u(x),\qquad \psi _{*}(x) = \lim _{c_{3}\to 0}
\psi (x),\qquad \wp _{*}(x) = \lim _{c_{3}\to 0}\wp (x). \]
By Lemma 12, the function $\wp _{*}(x)$ satisfies the equation
\begin{eqnarray*}
(\wp '_{*}(x))^{2}&=& 4\,\wp _{*}(x)^{3} - 3\left( \frac{2c_{2}}3\right) 
^{2}\wp_{*}(x) + \left( \frac{2c_{0}}{3}\right) ^{3}\\
&=& 4\left( \wp _{*}(x)-\frac{c_{2}}{3}\right) ^{2}\left( \wp _{*}(x)+
\frac23\,c_{2}\right) . \end{eqnarray*}
Therefore,
\begin{eqnarray}
(\psi '_{*}(x))^{2} &=&\frac14\left( \frac{-\wp '_{*}(x)}{(\wp _{*}(x)-
c_2/3)^{2}}\right) ^{2} \nonumber \\
&=&\frac{\wp _{*}(x)+2c_{2}/3}{(\wp _{*}(x)-c_{2}/3)^{2}} 
= 2\psi _{*}(x)+4c_{2}\,\psi _{*}(x). \end{eqnarray}
Differentiating (56) with respect to $x$, we obtain
\[ 
\psi _*''(x)=4\,c_2\psi _*(x)+1,\qquad \psi _*(0)=0,\quad \psi _*'(0)=0.\]
Therefore, 
\[
\psi _*(x)=\frac{\cosh 2\sqrt{c_2} x-1}{(2\,\sqrt{c_2})^2}. \]
In view of (51), 
\[ u_*(x)=c_1\,\psi _*(x)+c_0\,\psi _*'(x). \]
Corollary 13 is proved.

Note that according to Lemma 11, if the functions ($\varphi ,\tau ,A$) 
satisfy equation (34), then the function $\tau (x)$ is detrmined uniquely by 
the equation
\[
\frac{\tau'(x)}{\tau (x)}=\frac12\,\frac{u'(x)+c_0}{u(x)}\,,\]
subject to the initial conditions $\tau (0) = 0, \tau '(0) = 1$, and the
function $A(x)$ is determined by  equation (43):
\[
A(x) = \frac{u(x)}{\tau (x)}.\]
Hence, we may regard the functions $\varphi (x)$ as solutions of 
equation (34).

{\bf Theorem 14.} {\em The general solution of  equation} (34) 
\[
\varphi (x+y) = \varphi (x)+\varphi (y)+\tau (x)\tau (y)A(x+y) \]
{\em is given by the function}
\be
\varphi (x) = \frac{4}{c_{3}}\,(\zeta (\alpha )-\zeta (x+\alpha )-\wp
(\alpha )x),\qquad \varphi (0) = \varphi '(0) = 0, \ee
{\em where} $\zeta (x)$ {\em and} $\wp (x)$ {\em are the Weierstrass 
$\zeta $-function and $\wp $-function  with  parameters} $g_{2}$ 
{\em and} $g_{3}$ ({\em see Lemma} 12).

{\bf Proof.} According to Lemmas (11) and (12), it is sufficient to prove 
that any function $\varphi (x)$ given by  formula (57) is a solution of 
equation (14). It is convenient to consider two different cases.
\begin{description}
\item[{\bf Case 1.}] $c_{3} = 0.$
\[
\varphi _{*}(x) = \lim _{c_{3}\to 0}\varphi (x). \]
In this case, $\varphi _{*}(x) = \int _{0}^{\infty }u_{*}(x)dx$ and hence, 
using  Corollary 13, we obtain
\be
\varphi _{*}(x) = c_{1}\,\frac{\sinh 2\sqrt {c_{2}}x-2\sqrt {c_{2}}x}
{(2\sqrt {c_{2}})^{3}} + c_{0}\,\frac{\cosh 2\sqrt {c_{2}}x-1}
{(2\sqrt {c_{2}})^{2}}.\ee 
Using the elementary identity
\be 
e^{(x+y)}-1 = \left( e^{x}-1\right) + \left( e^{y}-1\right) + 
\left( e^{x/2}-e^{-x/2}\right) \left( e^{y/2}-e^{-y/2}\right) 
e^{(x+y)/2}, \ee
we obtain
\begin{eqnarray*}
\sinh 2\sqrt {c_{2}}(x+y) &=& \sinh 2\sqrt {c_{2}}x + \sinh 2\sqrt {c_{2}}y \\
&+& 4\sinh \sqrt {c_{2}}x \sinh \sqrt {c_{2}}y \sinh \sqrt {c_{2}}(x+y),\\
\cosh 2\sqrt {c_{2}}(x+y) &=& \cosh 2\sqrt {c_{2}}x + \cosh 2\sqrt {c_{2}}y \\
&+& 4\sinh \sqrt {c_{2}}x \sinh \sqrt {c_{2}}y \cosh \sqrt {c_{2}}(x+y).
\end{eqnarray*}
Hence,
\[
\varphi _{*} (x+y) = \varphi _{*}(x)+\varphi _{*}(y)+\tau _{*}(x)\tau _{*}
(y)A_{*}(x+y),\]
where
\be
\tau _{*}(x) = \frac{\sinh \sqrt {c_{2}}x}{\sqrt {c_{2}}},\qquad A_{*}(x) =
\frac{c_{1}}{2}\,\frac{\sinh \sqrt {c_{2}}x}{\sqrt {c_{2}}} + c_{0}\,
\cosh \sqrt {c_{2}}x.\ee

\item[{\bf Case 2.}] $c_{3} \neq 0$. Then without any restriction we may take
$c_{3} = 2$. According to the Frobenius--Stickelberger formula [FS 1880], 
the functions $f(x), g(y), h(z)$ constitute a solution of equation (1):
\begin{eqnarray}
f(x)&=& \zeta (\alpha _{1}-{\alpha }/{2}-x) - \wp (\alpha )x -\zeta
(\alpha _{1}-{\alpha }/{2}),\\
g(y) &=& \zeta (-\alpha _{1}-{\alpha }/2-y) - \wp (\alpha )y + \zeta 
(\alpha _{1}+{\alpha }/2),\\
h(z) &=& \zeta (\alpha -z) - \wp (\alpha )z - \zeta (\alpha ).
\end{eqnarray}
Using the reduction of (1) to equation (14) described above, we obtain 
\[
\varphi (x) = -\,2h(-x)=2\,(\zeta (\alpha )-\zeta (x+\alpha )-\wp 
(\alpha )x) \]
which gives the solution of equation (34). Theorem is proved.
\end{description}

{\bf Corollary 15.} {\em The general normalized solution of equation} (32) 
{\em is given by the formulas}
\begin{eqnarray}
\varphi (x) &=&\frac4{c_{3}}\,(\zeta (\alpha )-\zeta (x+\alpha )-\gamma
(\alpha )x),\nonumber \\
&&\\
\xi (x) &=& \frac{2u(x)}{c_{0}-2\,b_{3}u(x)+u'(x)},\nonumber \end{eqnarray}
{\em where}
\[
u(x) = \varphi '(x) =\frac4{c_{3}}\,(\wp (x+\alpha )-\wp (\alpha )) \]
{\em and} $b_{3}$ {\em is a free parameter,}
\[
\eta (x) = \varphi (x)-\varphi '(x)\xi (x),\qquad 
\gamma (x) = -\varphi '(x)\xi (x)^{2}.\]

The proof follows from Theorem 14, formula (61), and from Lemma 10. Let us 
recall that in the proof of Lemma 10 we gave an explicit construction 
of the solution to equation (12) using the solution of equation (14).

Thus, it is already proved that if $(f(x),g(y),h(z))$ is the 
nondegenerate solution of equation (1) satisfying the additional conditions
\be 
f(x) = g(x),\qquad f(0) = g(0) = h(0) = h'(0), \ee
then it is necessary  to have 
\be
h(x) =\frac2{c_{3}}\,(\zeta (\alpha -x)-\gamma (\alpha )x - \zeta
(\alpha )), \ee
where $c_{3}, \alpha $ and the parameters $g_{2},g_{3}$ of the 
$\wp $-Weierstrass function satisfy the condition of Lemma 12. Moreover, 
if $c_{3} \neq 0$, then for the functions
\begin{eqnarray}
f(x) &=& \frac2{c_{3}}\left( \zeta \left(\alpha _{1}-\frac{\alpha }{2}-x
\right) - \wp (\alpha )x - \zeta \left( \alpha _{1}-\frac{\alpha }{2}\right) 
\right) ,\\
g(x) &=& \frac{2}{c_{3}}\left( \zeta \left( -\alpha _{1}-\frac{\alpha }{2}
-x\right) - \wp (\alpha )x + \zeta \left( \alpha _{1}+\frac{\alpha }{2}
\right) \right) ,\end{eqnarray}
where $\alpha $ is the free parameter, the function $h(x)$ of form (66) 
gives the solution of equation (1). Hence, there are two unsolved problems.

\begin{description}
\item[{\bf 1.}] Are the functions $f(x)$ and $g(x)$ for $c_{3} \neq 0$ the 
only functions that give the solution of equation (1) for a fixed function 
$h(x)$?
\item[{\bf 2.}] How can we find sufficient conditions for $c_{3} = 0$ on 
the parameters of the function $h_{*}(x) = \lim _{c_{3}\to 0}h(x)$ such that 
there exist functions $f(x)$ and $g(x)$ for which $(f(x), g(x), h_{*}(x))$ is 
the solution of equation (1) and how can we find all such functions 
$(f(x), g(x))$?
\end{description}

Let us note that in the case $c_{3} = 0$ the main problem is that we cannot 
pass to the limit as $c_{3}\to 0$ in formulas (67), (68) (in contrast to 
(66)).

To solve these two problems we shall first consider the reduction of equation 
(1) to equation (12) and shall use the general analytic solution of 
equation (12) (see Lemma 9 and Corollary 15).

Let us begin with the case $c_{3} \neq 0$.

{\bf Lemma 16.} {\em Let the functions} $(f_{1}(x), g_{1}(x), h_{1}(x))$ 
{\em satisfy  equation} (1) {\em and the initial conditions under 
consideration. If} $h_{1}(x) = H(x)$ {\em is the function from equation} 
(66), {\em then}
\begin{eqnarray}
f_{1}(x) &=& s_{1}f(x)+s_{2}g(x),\\
g_{1}(x) &=& t_{1}f(x)+t_{2}g(x),\end{eqnarray}
{\em where} $f(x)$ {\em and} $g(x)$ {\em are given by equations} (67) 
{\em and} (68), {\em and} $s_{1}+s_{2} =1, t_{1}+t_{2} = 1$. 

{\bf Proof.} For the functions given by equations (47) and (48), we have 
\begin{eqnarray}
\xi (x) &=& f(x)-g(x)\nonumber \\
&=&\frac2{c_{3}}\left[ \zeta \left( \alpha _{1}-\frac{\alpha }{2}-x\right) 
+\zeta \left( \alpha _{1}+\frac{\alpha }{2}+x\right) \right.\nonumber \\
 &-&\left. \zeta \left( \alpha _{1}-\frac{\alpha }{2}\right) 
-\zeta \left( \alpha _{1}+\frac{\alpha }{2}\right) \right] .
\end{eqnarray}
Then
\begin{eqnarray*}
\xi '(x) &=&\frac2{c_{3}}\left[ \wp \left( \alpha _{1}-\frac{\alpha }{2}
-x\right) - \wp \left( \alpha _{1}+\frac{\alpha }{2}+x\right) \right] ,\\
\xi ''(x) &=&\frac{2}{c_{3}}\left[ -\wp '\left( \alpha _{1}-\frac
{\alpha }{2}-x\right) -\wp '\left( \alpha _{1}+\frac{\alpha }{2}+x\right) 
\right] . \end{eqnarray*}

We see that if the parameters $\alpha $ and $\alpha _{1}$ are sufficiently 
close to the point $x=0$, then $\xi '(0) \neq 0$, and the value $\xi ''(0)$ 
gives the value of free parameter $b_{3}$ required to construct the general 
normalized solution  of equation (32). Therefore, in this case the general 
solution of the equation has the form
\[
\varphi (x) = -2h(-x),\qquad \eta (x)+b_{1}\xi (x),\qquad b_{2}\xi (x), \]
where $h(x)$ is the function (66), $\xi (x)=f(x)-g(x)$ and $\eta (x)=
f(x)+g(x)$ for the functions (47) and (48).

Now if we introduce 
\[
f_{1}(x)+g_{1}(x) = \eta (x)+b_{1}\xi (x),\qquad 
f_{1}(x)-g_{1}(x) = b_{2}\xi (x), \]
then we get
\begin{eqnarray*}
f_{1}(x)&=&\frac{1}{2}\,\eta (x)+\frac{b_{1}+b_{2}}{2}\,\xi (x) = s_{1}f(x)+
s_{1}g(x),\\
g_{1}(x) &=&\frac12\,\eta (x)+\frac{b_{1}-b_{2}}{2}\,\xi (x) = t_{1}f(x)+
t_{2}g(x),\end{eqnarray*}
where
\[
s_{1} = \frac12+\frac{b_{1}+b_{2}}{2},\quad s_{2} =\frac12-\frac{b_{1}+
b_{2}}{2},\quad t_{1} =\frac12+\frac{b_{1}-b_{2}}{2},\quad t_{2} =
\frac12-\frac{b_{1}-b_{2}}{2}. \]
Lemma is proved.

Now it remains to find the values of parameters $s_{1}$ and $t_{1}$ for which 
the set of functions $(f_{1}(x), g(x), h(x))$ from  Lemma 16 gives the 
solution of equation (1).

Let us introduce the notation
\[
\det (f, g, h) =\mbox{det}\left( \begin{array}{ccc} 
f''(x)&g''(y)&h''(z)\\ f'(x)&g'(y)&h'(z)\\ 1&1&1 \end{array} \right) , \]
and use the following formula (see [WW 1927], p.458)
\[
\frac12\,\det (\wp (x),\wp (y),\wp (z)) = \frac{\sigma (x+y+z)\,\sigma 
(x-y)\,\sigma (y-z)\,\sigma (z-x)}{\sigma ^{3}(x)\,\sigma^{3}(y)\,
\sigma ^{3}(z)}.\]
If the conditions of Lemma 16 are satisfied, we have
\begin{eqnarray}
&&\det (s_{1}f(x)+s_{2}g(x),\,t_{1}f(x)+t_{2}\,g(x),\,h(z))\nonumber \\
&=&s_{1}t_{1}\,\det (f(x), f(y), h(z)) + s_{2}t_{2}\,\det (g(x), g(y),
h(z)) \end{eqnarray}
On the other hand,
\begin{eqnarray}
&&\frac{c_{3}^{3}}{8}\,\det (f(x), f(y), h(z))\nonumber \\
&=&\frac{c_{3}^{3}}{8}\,\det \left( \wp \left( \alpha _{1}-\frac{\alpha }{2}
-x\right),\quad \wp \left( \alpha _{1}-\frac{\alpha }{2}-y\right) ,\quad 
\wp (\alpha -z)\right) \nonumber \\
&=&\frac{c_{3}^{3}}{4}\,\frac{\sigma (2\alpha _{1})\sigma (y-x)\sigma (z-y+
\alpha _{1}-3\alpha /2)\,\sigma (x-z+3\alpha /2-\alpha _{1})}
{\sigma ^{3}(\alpha _{1}-\alpha /2-x)\,\sigma ^{3}(\alpha _{1}-
\alpha /2-y)\,\sigma ^{3}(\alpha -z)},\nonumber \\
&& \end{eqnarray}
\begin{eqnarray}
&&\frac{c_{3}^{3}}{8}\,\det (g(x), g(y), h(z)) \nonumber \\
&=&\frac{c_{3}}{8}\,\det (\wp \left( -\alpha _{1}-\frac{\alpha }2-x\right) ,
\wp \left( -\alpha _{1}-\frac{\alpha }2-y\right) ,\wp (\alpha _{1}-2)
\nonumber \\
&=&\frac{c_{3}^{3}}{4}\,\frac{\sigma (2\alpha _{1})\,\sigma (y-x)\,
\sigma (y-z+\alpha _{1}+3\alpha /2)\,\sigma (x-z+\alpha _{1}+3\alpha /2)}
{\sigma ^{3}(\alpha _{1}+\alpha /2+x)\,\sigma ^{3}(\alpha _{1}+
{\alpha }/2+y)\,\sigma ^{3}(\alpha - z)} . \nonumber \\
&& \end{eqnarray}

The comparison of expressions (73) and (74) shows that if $\alpha _{1} 
= \omega _{k}$ is the one of the three halfperiods of the 
Weierstrass-function $\wp (x)$, then det\,$(\cdot )$ given by formula (72) 
is equal to zero identically for all values of $s_{1}$ and $t_{1}$. 
However, if $\alpha _{1} \neq \omega _{k}, \,k =1,2,3$, then 
this determinant is equal to zero if and only if $s_{1}t_1= s_2t_{2} = 0$. 
So, we have proved our main result.

Now let us consider the case $c_{2} \to  0$.

In this case, the general normalized solution is given by function (58). 
Let us denote
\[
\varphi _{**}(x) = \lim _{c_{2}\to 0}\varphi _{*}(x),\qquad \tau _{**}(x) =
\lim _{c_{2}\to 0}\tau _{*}(x),\qquad A_{**}(x) = \lim _{c_{2}\to 0}A_{*}(x).
\]
From (58) we obtain
\be
\varphi _{**}(x) = c_{1}\,\frac{x^{3}}{{3!}}+c_{0}\,\frac{x^{2}}{2}. \ee
According to  formulas (40), we have
\[
\tau_{**}(x) = x,\qquad A_{**}(x) = c_{1}\,\frac{x}{2}+c_{0}.\]
Further,
\begin{eqnarray}
\xi _{**}(x) &=& \frac{x}{1-b_{3}x},\\
\eta _{**}(x) &=& c_{1}\,\frac{x^{3}}{{3!}}+c_{0}\,\frac{x^{2}}{2}-
\left( c_{1}\,\frac{x^{2}}{2}+c_{0}x\right) \frac{x}{1-b_{3}x},\\
\gamma _{**}(x) &=&-\left( c_1\frac{x^2}2+c_0x\right) \frac{x^2}{(1-b_3x)^2}.
\end{eqnarray}
Hence, in this case the general solution of equation (32) is given by 
the functions
\[
\varphi _{**}(x),\qquad \eta _{**}+b_{1}\xi _{**},\qquad b_{2}\xi_{**}.\]

{\bf Corollary 17.} {\em The general solution of equation} (14) {\em in the 
class of entire functions has the form}
\[ \begin{array}{ll}
f(x)=\alpha _1\,e^{\lambda x}+\beta x+\gamma _1, &\quad F(x)=(\alpha _1
e^{\lambda x}+\gamma/\sqrt{3})^2+2\alpha _2\alpha _3e^{-\lambda x},\\
g(x)=\alpha _2\,e^{\lambda x}+\beta x+\gamma _2, &\quad  G(x)=(\alpha _2
e^{\lambda x}+\gamma/\sqrt{3})^2+2\alpha _1\alpha _3e^{-\lambda x},\\
h(x)=\alpha _3\,e^{\lambda x}+\beta x+\gamma _3, &\quad H(x)=(\alpha _3
e^{\lambda x}+\gamma/\sqrt{3})^2+2\alpha _1\alpha _2e^{-\lambda x},
\end{array} \]
{\em where } $\gamma =\gamma _1+\gamma _2+\gamma _3$.

{\em For the case  in which} $\lambda \to 0$ {\em and for the corresponding} 
$\alpha _k$, $k=1,2,3$, $\beta $, {\em we obtain the solution}
\[ \begin{array}{ll}
f(x)=\alpha x^2+\beta _1x+\gamma _1, &\quad F(x)=2\alpha ^2(x-a_1)^4+2\alpha
\tilde \gamma (x-a_1)^2+\tilde \gamma ^2/3,\\
g(x)=\alpha x^2+\beta _2x+\gamma _2, &\quad G(x)=2\alpha ^2(x-a_2)^4+2\alpha
\tilde \gamma (x-a_2)^2+\tilde \gamma ^2/3,\\
h(x)=\alpha x^2+\beta _3x+\gamma _3, &\quad G(x)=2\alpha ^2(x-a_3)^4+2\alpha
\tilde \gamma (x-a_3)^2+\tilde \gamma ^2/3,\end{array} \]
{\em where}
\begin{eqnarray*}
a_1&=&\frac1{6\alpha }\,(\beta _1+\beta _3-2\beta _1),\qquad 
a_2=\frac1{6\alpha }\,(\beta _1+\beta _3-2\beta _2),\\
a_3&=&\frac1{6\alpha }\,(\beta _1+\beta _2-2\beta _3),\qquad 
\tilde \gamma =\gamma _1+\gamma _2+\gamma _3-\frac1{4\alpha }\,
(\beta _1^2+\beta _2^2+\beta _3^2).\end{eqnarray*}

\centerline{\bf Appendix}

It is interesting to note that the general solution of the functional
equation (1) has found the applications in another physical context. In 
[BB 1994] the Lax representation for the system of equations
\be
\ddot q_j=\sum _{k\neq j}(a+b\dot q_j)(q+b\dot q_k)\,V_{jk}(q_j-q_k),
\qquad j=1,\ldots ,n,\ee
was constructed. This system describes the motion of $n$ particles 
on the line. The particular cases of this system are integrable relativistic 
($b\neq 0$) and nonrelativistic ($b=0$) Calogero-Moser systems, 
as well as the Toda systems. Within the framework of this paper and 
following [BB 1994],  let us consider in more detail the case of three 
particles.

For the system of equations
\[
\ddot q_j=\sum _{k\neq j}V_{jk}(q_j-q_k),\qquad j=1,2,3, \]
let us search for the Lax representation $\dot L=[L,M]$ in the form
\[
L(q)=\dot q_d+A,\qquad M(q)=(B\tau )_d+C, \]
where $L,\,M,\,A,\,B,\,C$ are ($3\times 3$)-matrices and $A=(A_{jk}(q_j-
q_k))$, $A_{jj}\equiv 0$, $C=(C_{jk}(q_j-q_k))$, $C_{jj}\equiv 0$,
$B=(B_{jk}(q_j-q_k))$, $\tau =(1,1,1)$, and $\dot q_d$, $(B\tau )_d$ are
diagonal matrices whose diagonals contain the coordinates of the vectors
$\dot q=(\dot q_1,\dot q_2,\dot q_3)$ and $B\tau $, respectively.

The Lax representation leads to the equation
\be
\ddot q_d+[\dot q_d,A']=[A,(B\tau )_d]+[A,C]+[\dot q_d,C], \ee
where $A'=(A_{jk}'(q_j-q_k))$. Therefore,
\be 
\sum _{k\neq j} V_{jk}(q_j-q_k)=[A,C]_{jj},  \ee
\be
([\dot q_d,C-A']+[A,(B\tau )_d+C])_{jk}=0. \ee
From (82) we obtain $C=A'$ and hence, by (81),
\be
V_{jk}(q_j-q_k)=A_{jk}A_{kj}'-A_{jk}A_{kj}'=-V_{kj}(q_k-q_j). \ee
Further, (82) yields
\be
\sum _{l=1}^3 A_{jk}(B_{jl}-B_{kl})+A_{jl}'A_{lk}-A_{jl}A_{lk}'=0.\ee
Now let us set
\be
\Phi _{jk}=(A_{jl}'A_{lk}-A_{jl}A_{lk}')/A_{jk}. \ee
A direct verification shows that (84) implies the condition
\be
\Phi _{jk}+\Phi _{km}+\Phi _{mj}=0.\ee
Let us introduce the functions
\[ \begin{array}{l}
b_1(x)=-A_{23}(x)\,A_{32}(-x), \qquad b_2(y)=-A_{31}(y)\,A_{13}(-y),
\nonumber \\
b_3(z)=-\,A_{12}(z)\,A_{21}(-z).\nonumber  \end{array} \]
Taking $x=q_2-q_3$, $y=q_3-q_1$, $z=q_1-q_2$ and using the condition 
\[
 \Phi _{21}+\Phi _{13}+\Phi _{32}=0,\]
we immediately obtain
\be \begin{array}{l}
b_2(y)A_{23}A_{32}'-b_1(x)A_{31}A_{13}'+b_3(z)A_{31}A_{13}'\nonumber \\
-b_2(y)A_{12}A_{21}'+b_1(x)A_{12}A_{21}'-b_3(z)A_{23}A_{32}'=0.\nonumber 
\end{array} \ee
Similarly, from the condition
\[
 \Phi _{21}+\Phi _{23}+\Phi _{31}=0 \]
we get
\be \begin{array}{l}
b_1'(x)=A_{13}A_{31}'-b_2(y)A_{32}A_{23}'+b_3(z)A_{32}A_{23}'\nonumber \\
-b_1(x)A_{21}A_{12}'+b_2(y)A_{21}A_{12}'-b_3(z)A_{13}A_{31}'=0.\nonumber 
\end{array} \ee
Taking into account the relation
\be \begin{array}{l}
b_1'(x)= A_{23}A_{32}'-A_{32}A_{23}',\qquad b_2'(y)=A_{31}A_{13}'-
A_{13}A_{31}',\nonumber \\
b_3'(z)=A_{12}A_{21}'-A_{21}A_{12}',\nonumber \end{array} \ee
and adding equations (87) and (88), we obtain
\be
b_2(y)b_1'(x)-b_1(x)b_2'(y)+b_3(z)b_2'(y)-b_2(y)b_3'(z)+b_1(x)b_3'(z)-
b_3(z)b_1'(x)=0. \ee
Equation (90) may be rewritten as
\be 
 \mbox{det} \left( \begin{array}{ccc}b_1'(x)&b_2'(y)&b_3'(z)\\ 
 b_1(x)&b_2(y)&b_3(z)\\ 1&1& \end{array} \right) \equiv 0. \ee
Recall that in our notation $x+y+z=0$ and therefore, by Lemma 1, the 
functions 
\be
f(x)=\int b_1(x)dx,\qquad g(y)=\int b_2(y)dy,\qquad h(z)=\int b_3(z)dz \ee
satisfy equation (1) (which is identical to (14)) for the corresponding 
functions $F(x)$, $G(y)$, and $H(z)$.

Thus, we obtain the following result.

{\bf Theorem.} {\em The system of equations}
\[
\ddot q_j=\sum _{k\neq j}V_{jk}(q_j-q_k),\qquad j=1,2,3, \]
{\em has the Lax representation} $\dot L=[L,M]$ {\em of the form indicated 
above if and only if }
\be
V_{jk}(q_j-q_k)=-\,\alpha \wp '\,(q_j-q_k+\lambda _j-\lambda _k). \ee

{\bf Proof.} Suppose the system of equations does have the indicated Lax 
representation.  Then, by (83), (91), and (92),
\[
 V_{23}(x)=f''(x),\qquad V_{31}(y)=g''(y),\qquad V_{12}(z)=h''(z), \]
and therefore by the main theorem
\[
 V_{jk}(q_j-q_k)=-\,\alpha \rho '(q_j-q_k-q_l),\qquad l\not \in (j,k). \]
In view of condition (21), i.e., $a_1+a_2+a_3=0$, we see that  the $a_l$'s 
may be presented in the form $\lambda _j-\lambda _k$, which means that we 
have (92).

The proof of the converse statement is in fact contained in [Ca 1975b].


\bigskip
\begin{thebibliography}{POI}
\bibitem[Ab 1823]{PO} Abel N.H., {\em M\'ethode g\'en\'erale pour trouver 
des fonctions d'une seule quantit\'e variable, lorsqu'une propri\'et\'e de
ces fonctions est exprimee par une \'equation entre deux variables.}; 
Magazin for Naturvidenskaberne, Aargang I, Bind 1, Christiania; 
Oeuvres completes, {\bf 1}, Christiania, (1881), 1-10

\bibitem[BB 1994]{KK} Braden H.W. and Buchstaber V.M., {\em Integrable 
systems with pairwise interactions and functional equations}, Preprint, 
hep-th/9411240 

\bibitem[BC 1990]{KJ} Bruschi M. and Calogero F., {\em General analytic 
solution of certain functional equations of addition type}, SIAM J. Math. 
Anal. {\bf 21} 1019--1030

\bibitem[BFV 1994]{GF} Buchstaber V., Felder G. and Veselov A., {\em 
Elliptic Dunkl operators, root systems and functional equations}, Duke 
Math. J. {\bf 76}  No.3, 385--911

\bibitem[BK 1993]{BV} Buchstaber V. and Krichever I., {\em Vector addition 
theorems and Baker-Akhiezer functions}, Teoret. Mat. Fiz. {\bf 94}, No.2, 
200--212; English transl. in Theor. and Math. Phys. (1993)

\bibitem[BP 1993]{NB} Buchstaber V. and Perelomov A., {\em On the functional 
equation related to the quantum three-body problem}, Preprint MPI/93--17 

\bibitem[Ca 1821]{NB} Cauchy A.L., {\em Cours d'Analyse de l'Ecole Polyt.}, 
{\bf 1}; {\em Analyse algebraique,} {\bf 103}, Oeuvres completes (2) {\bf 3}, 
(1821), 98-105

\bibitem[Ca 1975a]{CC} Calogero F., {\em One-dimensional many-body problems 
with pair interactions whose exact ground-state wave function is of product 
type}, Lett. Nuovo Cimento {\bf 13} 507-511

\bibitem[Ca 1975b]{MM} Calogero F., {\em Exactly solvable one-dimensional 
many-body problems}, Lett. Nuovo Cimento (2) {\bf 13} 411-416

\bibitem[FS 1880]{CX} Frobenius G. and Stickelberger L., {\em \"Uber die 
Addition und Multiplication der elliptischen Functionen}, J. Reine Angew. 
Math. {\bf 88}  146-184

\bibitem[OP 1983]{DS} Olshanetsky M.A. and Perelomov A.M., {\em Quantum 
integrable systems related to Lie algebras}, Phys. Rep. {\bf 94} 
313-404

\bibitem[Pe 1990]{MN} Perelomov A.M. {\em Integrable Systems of Classical 
Mechanics and Lie Algebras}, Birkhauser 

\bibitem[Su 1975]{VC} Sutherland B., {\em Exact ground-state wave function 
for a one-dimensional plasma}, Phys. Rev. Lett. {\bf 34}  1083-1085

\bibitem[WW 1927]{NB} Whittaker E. and Watson G., {\em Course of Modern 
Analysis}, Cambridge Univ. Press 

\end{thebibliography}
\end{document}